\documentstyle[12pt]{report}

\newcommand{\semi}{;\hfil\break}
\newcommand{\e}[1]{\label{eq:#1}}
\newcommand{\ee}[1]{(\ref{eq:#1})}

\newcommand{\eq}{\begin{equation}}
\newcommand{\eqe}{\end{equation}}
\newcommand{\eqa}{\begin{eqnarray}}
\newcommand{\eqae}{\end{eqnarray}}

\newcommand{\del}{\partial}
\newcommand{\zb}{{\bar{z}}}
\newcommand{\blambda}{\bar\lambda}

\newcommand{\bra}[1]{\mbox{$\langle #1 $}}
\newcommand{\ket}[1]{\mbox{$| #1 \rangle$}}

\begin{document}

\pagestyle{empty}
\hfill{NSF-ITP-93-62}

\hfill{UCSBTH-93-14}

\hfill{hep-th/9305083}

\vspace{18pt}

\begin{center}
{\large \bf Four Dimensional Black Holes in String Theory}

\vspace{24pt}
Steven B. Giddings\\[6pt]
Joseph Polchinski\\[6pt]
Andrew Strominger\\

\vspace{12pt}
\sl

Institute for Theoretical Physics \\ University of California \\
Santa Barbara, California 93106-4030

\vspace{12pt}
\rm and\\
\vspace{12pt}
\sl
Department of Physics \\ University of California
\\ Santa Barbara, CA 93106-9530
\rm

\vspace{12pt}
{\bf ABSTRACT}

\end{center}

\begin{minipage}{4.8in}
Exact solutions of heterotic string theory corresponding to
four-dimensional charge $Q$ magnetic black holes are constructed as tensor
products of an $ SU(2)/Z(2Q+2)$ WZW orbifold with a $(0,1)$ supersymmetric
$SU(1,1)/U(1)$ WZW coset model.  The spectrum
is analyzed in some detail.
``Bad'' marginal operators are found which
are argued to deform these theories to asymptotically flat black holes.
Surprising behaviour is found for small values of $Q$, where low-energy
field theory is
inapplicable. At the minimal value
$Q=1$, the theory degenerates. Renormalization group arguments are given that
suggest the potential gravitational singularity of the
low-energy field theory is resolved by a massive two-dimensional field theory.
At $Q=0$, a stable, neutral ``remnant,'' of
potential relevance to the
black hole information paradox, is found.
\end{minipage}

\vfill

\pagebreak
\pagestyle{plain}
\setcounter{page}{1}
\baselineskip=16pt
\section*{1.\ \ Introduction}
\refstepcounter{chapter}

One of our expectations of string theory is that as a quantum theory of
gravity it should help us understand the puzzles of black holes, possibly
by resolving the problem of curvature singularities.
Perhaps a complete understanding of this must await a fully
non-perturbative understand of the theory, but we have begun to investigate
the role of singularities by studying exact classical string solutions.  In
particular, in \cite{Witt} Witten found an exact conformal field theory
corresponding to a two-dimensional dilatonic black hole.  Although the
stringy meaning of the singularity in this solution is not fully
understood, it does serve as a well-defined example in which to address the
questions.  One concern, however, is that this two-dimensional example is
oversimplified:  singularities in higher-dimensional black holes may be
qualitatively different.  For this reason classical string solutions
corresponding to four-dimensional black holes should also be useful.  Such
solutions have previously been found
in \cite{GiMa,GHS} as solutions of the low-energy effective theory
for string theory.  Although these charged solutions are known to leading
order in the $\alpha'$ expansion, corresponding exact solutions had not
yet been found.

In this paper exact string solutions
corresponding to certain limiting cases
(in which the asymptotic two spheres have finite radius) of
the
magnetic black holes of ref.~\cite{GiMa,GHS} will be given.
In these limiting cases the solutions become simple
products of the black hole of \cite{Witt} and a non-singular conformal
field theory on a two sphere.  The singularities found
are therefore identical to those of the two-dimensional black holes.
Further, the general solutions are
expected to arise from these limiting cases by a deformation corresponding to
a (1,1) operator in the conformal field theory that we will construct.
This suggests that even for the general solution the singularity is closely
related to the two-dimensional version.
Finally, it should be noted that such solutions may be relevant to the
real world, if it is described by string theory.  This is seen by noting
that for black holes whose radii are small compared to the dilaton Compton
wavelength the dilaton is effectively massless.  This means that real black
holes below this scale could be described by the string solutions of this
paper\cite{HoHo,GrHa,Horo}.

An important virtue of having the exact solutions is that it
will allow us to study small values of the monopole
charge, outside of the validity of the $\alpha'$ expansion.
We find some surprises
for small values of the charge, namely two solutions with unexpected
properties.

In the $\alpha'$ expansion the balance between magnetic field
and curvature gives the throat a radius proportional to the monopole charge,
$r_2 = Q$.
In particular, there would not seem to be a neutral
solution.  In the exact solution this is corrected to $r_2 =
|Q^2 - 1|^{1/2}$, and so a $Q=0$ solution of finite radius exists.
There is an extremal $Q=0$ solution with zero Hawking temperature,
$\it i.e.$ a neutral remnant. This is of obvious interest to
advocates (among which we do not necessarily count ourselves!) of neutral
remnants as a solution to the black hole information paradox.

Equally surprising is the fact that at
$|Q| = 1$
the radius is zero.  The solution is trivial in the sense of conformal
field theory, but that need not mean that it is uninteresting!
There do not yet exist methods to analyze the interpolation between the
exact throat and flat exterior solutions; this is one of the important
open questions in this work.  We will
introduce an approximate picture, based on the renormalization group,
which suggests that there exist solutions where the $Q=0, \pm 1$ throats
do connect onto the flat exterior.
In particular, at $|Q| = 1$, the infinite
throat narrows steadily to zero radius.  This is perhaps a counterexample
to the phenomenon of duality observed in toroidal compactification: down
the throat, almost all states of the string move to arbitrarily high mass.

In this conjectured form of the $|Q|=1$ solution, the would-be singularity
is resolved by a massive two-dimensional field theory.
The nature of the singularity is essentially stringy, and
does not have a spacetime interpretation. The generality of our arguments
invite the speculation that this
is a general method for resolving singularities in string theory.

The solutions factor into an $xt$ CFT, an angular CFT and an internal CFT.
The $xt$ CFT is the linear dilaton or two-dimensional black hole,
already known exactly.  The angular `monopole'
CFT has a central charge which approaches
3 in the semiclassical limit. Rotational invariance implies the existence of
an $SU(2)$ current algebra. Both of these observations point to
a close connection with the WZW
sigma model.  In fact, we will find that the monopole theory is a coset of
the level $k = 2|Q^2 - 1|$ $SU(2)$
WZW model by a discrete subgroup.  For
$Q > 1$ this is
\eq
\frac{SU(2)_L \times SU(2)_R}{Z(2Q+2)_R};  \e{co}
\eqe
to make a $\hat c = 4$ CFT the two-dimensional black hole factor would be
the level $k' = 2(Q^2 + 1)$, $SU(1,1)/U(1)$ coset theory.

In section 2 we
review the low-energy effective field theory
description of the magnetic black holes,
and present the heterotic sigma model. In
section 3 we bosonize the action and find that
the result is indeed the coset~\ee{co}
of the WZW model.  We also describe a mild
generalization, a bosonic theory with independent left- and right-moving
monopole charges $Q_{L,R}$.  In section~4
we bosonize the vertex operators.  Taking appropriate monopole harmonic
wavefunctions, the vertex operators are WZW vertex operators with a twist
of the right-moving $U(1)$ boson.  In section~5 we take a more abstract
approach, looking for consistent level-matched string
theories of the form~\ee{co}, and recover
the monopole theories constructed earlier.

In section~6
we return to the heterotic case, where $Q_R$ takes the value~1
appropriate for the spin connection.  We verify world-sheet supersymmetry
(which is $(0,2)$ due to a $U(1)$ symmetry), check level matching, and show
that there is no spacetime
supersymmetry.  We show that the solutions are in general
unstable if the monopole $U(1)$ is embedded in a
non-Abelian group (as expected from low-energy considerations), but
can be stable otherwise.  We find the vertex operator corresponding to
a widening of the throat toward the mouth, but
are unable to extend the exact solution through the mouth to the asymptotic
spacetime.  Also, we verify that the exact solution
is consistent with the index theorem for massless
fermions in a monopole field.  In section~7 we present the $Q=0$ solution,
which is a slight variation of the $|Q| > 1$ case.  In section~8 we discuss
$|Q| = 1$.  Our main tool is an approximate identification
of the radial dependence of the solution with a renormalization group
flow.  The monopole CFT becomes strongly coupled and develops a mass gap in
the region of the mouth, leaving a trivial CFT in the throat.  The index
theorem is useful in
understanding the physics of the solution, and world-sheet
instantons play an essential role.
One of the main open problems is to
obtain more control over the mouth region and verify that the existence of a
solution interpolating between the known throat and asymptotic theories.

It is worth mentioning that
the solution might have been presented
in a different - and much shorter -
manner. One could simply demonstrate that
the coset model~\ee{co} obeys all the criteria for a
building block of a consistent
heterotic string theory: modular invariance, $(0,1)$ supersymmetry
and a suitable GSO projection.  The identification of~\ee{co}
as a charge $Q$ monopole on $S_2$ then follows from the existence
of $SU(2)$ rotational symmetries together  with $2Q$
massless spacetime fermions. We have instead chosen the scenic route to the
final result, along which one views in detail the beautiful interplay
between the current algebra, spacetime and sigma
model descriptions of the theory.  The insights gained via this route
are important in our interpretation of the unexpected phenomona
at $Q=0$ and $Q=1$.

There is by now an extensive literature on stringy black holes.  An
excellent review with references can be found in ref.~\cite{Horo}.

\section*{2.\ \ Review of Low-Energy Solutions}
\refstepcounter{chapter}

\def\roughly#1{\raise.3ex\hbox{$#1$\kern-.75em\lower1ex\hbox{$\sim$}}}
\def\half{\frac{1}{2}}

The exact heterotic string solution that we will construct corresponds to the
extremal member of a family of magnetically charged four-dimensional
dilatonic black holes.
These black holes first appeared\cite{GiMa,GHS} as solutions of the
the low-energy effective action
\eq
S_4 \propto
\int d^4 x \sqrt{-g}
e^{-2 \Phi} \left(
R + 4(\nabla \Phi)^2 - {\alpha' \over 8} tr(F^2)  \right)
\eqe
which results when strings are compactified to four dimensions.
We henceforth use units in which
$\alpha' = 2$.
The black hole solutions are parametrized by the
sigma-model mass $M$, the charge $Q$,
and the value of the dilaton at infinity, $\Phi_0$.
For general values of these parameters
they take the form
\eqa
ds^2 &=& -4Q^2 \tanh^2 \sigma dt^2
 +\Bigl( 2M + \Delta\sinh^2\sigma \Bigr)^2
 \Bigl( 4d\sigma^2 + d\Omega_2^2 \Bigr) \nonumber\\
e^{2(\Phi-\Phi_0)} &=& \frac{2M + \Delta \sinh^2\sigma
}{\Delta \cosh^2 \sigma} \nonumber\\
F  &=& Q \epsilon_2\ ,  \e{solution}
\eqae
where $d\Omega_2^2$ and $\epsilon_2$ are the line element
and volume form on the unit two sphere, and we use a $U(1)$ embedding
with ${\rm Tr}(F^2) = 2F_{\mu\nu} F^{\mu\nu}$.
Here
\eq
\Delta=2M-{Q^2 \over 2M}.
\eqe
To give a complete string solution
eq.~\ee{solution} should be supplemented by an
internal solution corresponding to the compactification to four
dimensions.

The spatial geometry of a constant $t$ slice through this solution is shown
in Figure~1.  In the limit $M\rightarrow Q/2 $ the throat length
approaches infinity as $Q \ln\left(Q/\Delta\right)$.  Near this
limit there are four distinct regions,
\eq
\begin{array}{lll}
{\rm i)}\qquad\qquad\qquad &\sigma\gg \half \ln (Q/\Delta)
&{\rm asymptotically\ flat\ region}\cr
{\rm ii)} &\sigma\sim  \half \ln (Q/\Delta) & {\rm mouth}\cr
{\rm iii)} &\half \ln (Q/\Delta)\gg \sigma \gg 1\quad   & {\rm throat}\cr
{\rm iv)}& \sigma\roughly<1 & {\rm black\ hole}.
\end{array}
\eqe
At the limit $M= Q/2 $ one then finds three distinct solution depending on
where
one's attention is focussed and how the dilaton
is held fixed while taking the limit.  The
asymptotically flat region plus infinite throat is
\eqa
ds^2 &=& -4 Q^2dt^2 +\left(1+\frac{Q}{ y}
\right)^2  \left(dy^2 +y^2 d \Omega^2\right)\ , \nonumber\\
e^{2(\Phi-\Phi_0)} &=& 1+\frac{Q}{ y}\ ,\nonumber\\
F &=& Q \epsilon_2\
\eqae
where $y = \Delta \cosh^2\sigma$.
If instead $\sigma=x+\sigma_0$ where
$\half\ln (Q/\Delta)\gg \sigma_0 \gg 1$
the $M= Q/2 $ limit gives the throat solution:
\eqa
ds^2 &=& -4Q^2dt^2 +4Q^2dx^2 + Q^2
d\Omega_2^2,\nonumber\\
\Phi &=& -x+ {\tilde \Phi}_0,  \nonumber\\
F &=& Q \epsilon_2\ .  \e{that}
\eqae
Here the additive constant in the dilaton is shifted so that the dilaton is
finite at finite $x$ rather than at infinity.
Finally the black hole plus infinite throat is given by
\eqa
ds^2 &=& -4Q^2 \tanh^2 \sigma
dt^2 +  4Q^2 d\sigma^2 +
Q^2 d\Omega_2^2\ ,\nonumber\\
e^{2(\Phi-\hat\Phi_0)} &=& \frac{ Q}{ \cosh^2 \sigma }\ ,\nonumber\\
F &=& Q \epsilon_2\   \e{this}
\eqae
where once again the dilaton is shifted to be finite in the region of
interest.
Since the asymptotically flat region has disappeared
in the latter two
limits, it is not appropriate to associate the mass $M= Q/2 $
with the spacetime. An application of
the ADM procedure to spacetimes
of the form \ee{this} which are asymptotic to \ee{that}
yields a mass proportional to $e^{2\hat\Phi_0}$\cite{Witt}.

In both of these latter cases the solutions are trivial products of two
two-dimensional solutions.  In both the angular solution is the round
two-sphere with constant radius
threaded by a uniform magnetic flux.  The other solution is in
the first case the linear dilaton together with time, and in the second the
two-dimensional black hole of \cite{MSW,Witt}.  Both
the linear dilaton vacuum and the black hole correspond to exact conformal
field theories.  A similar factorization occurs for analogous
five-dimensional black holes in the extremal limit, with $S^2$ replaced by
$S^3$ with torsion.  In that case the angular theory corresponds to an
exact conformal field theory and therefore the five-dimensional
solution is an exact string solution\cite{wsheets,exbf}
when supplemented by an extra five dimensions.
This together with the simplicity of the above
$S^2$ theory leads to the conjecture that it also corresponds to
an exact conformal field theory, and thus yields an exact string solution
representing a four-dimensional black hole.
The present paper will construct that theory.

A world-sheet description of such a solution is via a heterotic sigma model,
with action\footnote
{Our conventions are $\alpha' = 2$ and $d^2 z = 2
d\sigma d\tau$.  The current algebral fermion $\lambda_L$ has charge $e = 1$.
The action is well defined for $Q$ half-integer, but we will find below that
$Q$ must in fact be an integer.}
\eqa
S &=& {1\over4\pi} \int d^2z\, r_2^2 G^{S_2}_{\mu\nu}
\del_z X^\mu \del_\zb X^\nu
+ {1\over2\pi} \int d^2z\,
\Biggl\{ \bar\lambda_R (\del_\zb - i \omega_\mu \del_\zb X^\mu ) \lambda_R
\nonumber\\
&&\qquad\qquad\qquad
\blambda_L (\del_z - i 2Q A_\mu^M \del_z X^\mu) \lambda_L +
i2Q F^M_{\mu\nu} \psi_R^\mu \psi_R^\nu \blambda_L \lambda_L
\Biggr\}\ .
\eqae
Here $G_{\mu\nu}^{S^2}$ is the unit round metric on the two sphere,
$\lambda_L$ is the current algebra fermion,
$\lambda_R$ is the supersymmetric fermion in tangent space,
\eq
\lambda_R = (e_\mu^1 + ie_\mu^2)\psi_R^\mu\ ,
\eqe
$\omega_\mu$ is the spin connection for vectors on the two sphere,
and $A_\mu^M$ is the gauge field of a magnetic monopole of unit
charge.
In the `northern'
coordinate patch the monopole potential is
\eq
A_\phi^{M(N)} = \frac{1 - \cos \theta}{2},
\eqe
and in the `southern' patch it is
\eq
A_\phi^{M(S)} = A_\phi^{M(N)} + i e^{-i\phi} \del_\phi e^{i\phi}
= - \frac{1 + \cos \theta}{2}.
\eqe

The four fermion interaction is necessary for world-sheet
supersymmetry.
The one-loop beta-function equations fix
the radius $r_2$ in terms of the charge $Q$,
\eq
r_2^2 = Q^2\ .
\eqe
The spin connection on the two-sphere
is simply a monopole field of charge~1, and the four-fermi interaction
can be rewritten using
\eqa
F^M_{\mu\nu} \psi_R^\mu \psi_R^\nu
&=& \frac{1}{2r_2^2} \epsilon_{\mu\nu} \psi_R^\mu \psi_R^\nu
 \nonumber\\[2pt]
&=& -\frac{i}{2r_2^2}  \bar\lambda_R \lambda_R.
\eqae
The action then takes the form
\eqa
S \!&=&\! \frac{1}{4\pi} \int d^2z\, (G_{\mu\nu} + B_{\mu\nu})
\del_z X^\mu \del_\zb X^\nu
+ \frac{1}{2\pi} \int d^2z\,
\Biggl\{ \blambda_R (\del_\zb - i A_{R\mu} \del_\zb X^\mu ) \lambda_R
\nonumber\\
&&\qquad\qquad\qquad +
\blambda_L (\del_z - i A_{L\mu} \del_z X^\mu) \lambda_L -
h \blambda_R \lambda_R \blambda_L \lambda_L \Biggr\}  \e{fermact}
\eqae
with backgrounds
\eqa
G_{\mu\nu} &=& r_2^2 G_{\mu\nu}^{S_2} \nonumber\\[2pt]
B_{\mu\nu} &=& 0 \nonumber\\[2pt]
A_{R\mu} &=& 2Q_R A_\mu^{M} \nonumber\\[2pt]
A_{L\mu} &=& 2Q_L A_\mu^{M}
\eqae
and
\eq
Q_L = Q, \qquad Q_R = 1,
\eqe
and a Thirring coupling, $h = -Q /
r_2^2$.
The world-sheet fermions in the north and south patches are related by
\eq
\lambda_{R,L}^{(S)} = e^{-i 2 Q_{R,L} \phi} \lambda_{R,L}^{(N)},
\e{lampatch}
\eqe
In order to make connection with the WZW model, we will
bosonize the left- and right-moving fermons.  The worldsheet
supersymmetry will then not be explicit, and it will be just as easy
to work with general charges $Q_{L,R}$.  In section~6 we will return to
the supersymmetric case $Q_R = 1$.

\section*{3.\ \ Equivalence to WZW Action}
\refstepcounter{chapter}

We now bosonize~\ee{fermact}, introducing a third embedding field $X^3$
which is periodic with period $2\pi$.  The bosonized action is
\eqa
S \!&=&\! \frac{1}{4\pi} \int d^2z\, \Biggl\{  r_2^2 G^{S_2}_{\mu\nu}
\del_z X^\mu \del_\zb X^\nu + r_1^2 (\del_z X^3 - A_\mu^+ \del_z X^\mu)
(\del_\zb X^3 - A_\mu^+ \del_\zb X^\mu)
\nonumber\\
&& +
A_\mu^- (\del_z X^3 \del_\zb X^\mu - \del_\zb X^3 \del_z X^\mu)
+\frac{1}{2} A_\mu^- A_\nu^+ (\del_z X^\mu \del_\zb X^\nu -
\del_z X^\nu \del_\zb X^\mu)
\Biggr\},\nonumber\\[4pt] \e{3sig}
\eqae
where
\eq
r_1^2 = 1 + 2h
\eqe
and
\eq
A_\mu^{\pm} = A_\mu^L \pm A_\mu^R = 2Q_\pm A_\mu^M, \qquad Q_{\pm} = Q_L \pm
Q_R. \e{back}
\eqe
The invariances
\eqa
X^3 &\to& X^3 + \epsilon^L (X^\mu) + \epsilon^R(X^\mu) \nonumber\\[2pt]
A_\mu^L &\to& A_\mu^L + \del_\mu \epsilon^L  \nonumber\\[2pt]
A_\mu^R &\to& A_\mu^R + \del_\mu \epsilon^R  \nonumber\\[2pt]
B_{\mu\nu} &\to& B_{\mu\nu} + \epsilon^L F^L_{\mu\nu} -
\epsilon^R F^R_{\mu\nu}  \e{eplus}
\eqae
provide a check that the gauge field has been correctly
introduced.\footnote
{The last term in the action vanishes for the particular
background~\ee{back}, but was included to allow independent $L$ and $R$
gauge transformations.}
Notice in particular that the $A_\mu^-$ term must have an $h$-independent
coefficient because it corresponds to torsion, which is quantized.
Similarly, the two terms in $\del_z X^3 + A_\mu^+ \del_z X^\mu$ must have
$h$-independent relative coefficient in order that the mapping between the
north and south coordinate patches respect the $2\pi$ periodicity of $X^3$.

The quantization of the torsion is seen explicitly if one integrates
by parts,
\eq
- \frac{1}{4\pi} \int d^2z\,
X^3 F_{\mu\nu}^- \del_z X^\mu \del_\zb X^\nu.
\eqe
Under $X^3 \to X^3 + 2\pi$, this changes by
\eq
-\frac{i}{4} \int F^-_{\mu\nu} dX^\mu dX^\nu = -i 2\pi nQ_-,
\eqe
where $n$ is the winding number of the map from the world-sheet to
$S_2$.
The path integral is therefore well-defined for all $n$ if
$Q_-$ is integer.  The restriction to integer values is due to a global
world-sheet anomaly.  In the fermionic language, if we take a genus zero
world-sheet mapped once to the spacetime two-sphere, there are $2Q_-$ net
fermionic zero modes; this number must be even.  There is also a spacetime
interpretation: in the left-moving Ramond sector, the charges are
half-integer.

The three-dimensional target space of the bosonized theory has curvatures
\eqa
R_{\hat 1 \hat 1} &=& R_{\hat 2 \hat 2} \ =\ \frac{2r_2^2 - r_1^2
Q_+^2}{2 r_2^4} \nonumber\\[2pt]
R_{\hat 3 \hat 3} &=& \frac{r_1^2 Q_+^2}{2 r_2^4}
\eqae
and torsion
\eq
H_{\hat 1 \hat 2 \hat 3} = -Q_-/r_2^2 r_1,
\eqe
where we have for convenience used tangent space indices
\eq
e_{\hat \imath} = \Bigl( r_2 d\theta, r_2 \sin \theta d\phi, r_1 (dX^3 -
A^+_\phi d\phi) \Bigr).
\eqe
The one-loop beta functions are then
\eqa
\mu \del_\mu \ln r_2 &=& R_{\hat 1 \hat 1} - \frac{1}{2} H^2_{\hat 1 \hat 2
\hat 3} \nonumber\\
&=& \frac{2 r_2^2 r_1^2 - r_1^4 Q_+^2 - Q_-^2}{2 r_2^4 r_1^2}
\nonumber\\[8pt]
\mu \del_\mu \ln r_1 &=& R_{\hat 3 \hat 3} - \frac{1}{2} H^2_{\hat 1 \hat 2
\hat 3} \nonumber\\
&=& \frac{r_1^4 Q_+^2 - Q_-^2}{2 r_2^4 r_1^2}.
\eqae
By redefining $\lambda_{R} \leftrightarrow \blambda_{R}$,
$\lambda_{L}\leftrightarrow \blambda_{L}$, and $z \leftrightarrow \zb$,
we may assume
\eq
Q_L \geq Q_R \geq 0.
\eqe
The one-loop beta functions then vanish for
\eq
r_2 = \sqrt{Q_+ Q_-},\qquad r_1 = \frac{ r_2}{Q_+}
= \sqrt{\frac{Q_-}{Q_+}}. \e{r1r2}
\eqe
This also implies for the Thirring coupling
\eq
h = -\frac{Q_R}{Q_+}. \e{gthir}
\eqe

The action is now
\eqa
S &=& \frac{ Q_+ Q_-}{4\pi} \int d^2z\, \Biggl\{ (G^{S_2}_{\mu\nu}
+ 4 A_\mu^M A_\nu^M)
\del_z X^\mu \del_\zb X^\nu
\nonumber\\
&&\qquad\qquad\qquad\qquad
+ \frac{1}{Q_+^2}
(\del_z X^3 - 4 Q_+ A_\mu^M \del_z X^\mu) \del_\zb X^3
\Biggr\}. \e{sigsol}
\eqae
As we have discussed in the introduction, one expects a close relation to
the SU(2) WZW model\cite{WZW}.
Topologically, the sigma model~\ee{sigsol} is an
$S_1$ bundle over $S_2$, with winding number $2Q_+$ and torsion
$\frac{1}{8\pi^2} \int H = Q_-$.   The $SU(2)$ group
manifold is also an $S_1$ bundle over $S_2$ (the Hopf fibration).
Writing the group element in terms of Euler angles
\eq
g = e^{i \phi \sigma_3/2} e^{i \theta \sigma_2/2} e^{i(\xi - \phi)
\sigma_3/2} \e{euler}
\eqe
with ranges
\eq
0 \leq \theta \leq \pi, \qquad 0 \leq \phi \leq 2\pi,
\qquad 0 \leq \xi \leq
4\pi,
\eqe
the fiber coordinate is $\xi$ and the base coordinates $\theta,\phi$.
One sees from eq.~\ee{euler} that at the north pole ($\theta = 0)$,
$\xi$ is a good coordinate, while at the south pole ($\theta = \pi)$, $\xi
- 2\phi$ is a good coordinate.  Noticing the ranges, the winding number is
one.  Thus, if we have an SU(2) WZW model of torsion (level)
\eq
k = 2 Q_+ Q_-,
\eqe
and if we identify
\eq
\xi = \frac{1}{Q_+} X^3, \qquad
\xi \sim \xi + \frac{2\pi}{Q_+},  \e{period}
\eqe
the group manifold is
is topologically the same as the sigma model~\ee{sigsol}.  Not
surprisingly, one finds that in terms of Euler angles the WZW action\cite{WZW}
is
precisely eq.~\ee{sigsol}.
The exact central charge is then
\eq
c = \frac{3k}{k + 2}.
\eqe

The monopole model~\ee{fermact} is invariant under $SU(2)$ spatial
rotations and under one non-anomalous linear combination of chiral fermion
rotations, $Q_L\epsilon_L = Q_R \epsilon_R$. Correspondingly, the WZW model
with identification~\ee{period} is invariant under $SU(2)$
left-multiplication and $U(1)$ right-multiplication, the remainder of the
right $SU(2)$ being inconsistent with the identification.  The Noether
current of the $U(1)$, $\xi \to \xi + \epsilon$, is
\eqa
j_{Nz} &=& \frac{i Q_-}{2} (\del_z X^3 - 4 Q_+ A_\mu^M \del_z X^\mu)
\nonumber\\[2pt]
j_{N\zb} &=& \frac{i Q_-}{2} \del_\zb X^3  \e{curr1}
\eqae
This current is not chiral, but becomes so after a trivial redefinition
$(j_z, j_\zb) = (j_{Nz}, j_{N\zb}) + i Q_- (\del_z X^3, -\del_\zb
X^3)/2$,
\eqa
j_z &=& i Q_- (\del_z X^3 - 2 Q_+ A_\mu^M \del_z X^\mu)
\ =\ \frac{k}{2}\ {\rm tr}(\sigma^3 g^{-1} \del_z g)
\nonumber\\[2pt]
j_\zb &=& 0. \e{curr2}
\eqae
{}From the $SU(2)$ current algebra, we have
\eq
j_z(z) j_z(0) \sim \frac{k}{2z^2}.
\eqe
We can write $j_z$ as the gradient of an analytic scalar,
\eq
j_z(z) = \frac{i k}{2} \del_z \xi_R,
\eqe
with
\eq
\xi_R(z) \xi_R(0) \sim -\frac{2}{k} \ln z.  \e{xixi}
\eqe
The normalization is fixed so that $j_z(z) \xi_R(0) = -i/z$.
Notice that $\xi_L = \xi - \xi_R$ is not
antianalytic, and $\del_\zb \xi_L$ is not a conserved current.

In summary, the monopole CFT is
\eq
\frac{SU(2)_L \times SU(2)_R}{Z(2Q_+)_R}. \e{coset}
\eqe
with the unbroken $SU(2)$ currents being antianalytic and the
unbroken $U(1)$ being analytic.  In our conventions right-moving (analytic)
happens to coincide with right-multiplication.
The $z$-$\zb$ asymmetry of the construction arises from the inequality $Q_R
< Q_L$. For $Q_R = Q_L$, the level $k$ is zero and the CFT is trivial; we
will discuss this case further in section 8.

\section*{4.\ \ The Spectrum}
\refstepcounter{chapter}

To complete the identification, we will express the vertex operators of
the monopole theory in terms of the WZW operators.  The fermions
$\lambda_{R,L}$ have momenta $p_3 = \frac{1}{2}$ and winding numbers $w_3 =
\pm 1$ in the 3-direction.  The right- and left- fermion numbers are then
\eq
F_R = p_3 + \frac{w_3}{2},\qquad F_L = p_3 - \frac{w_3}{2}.
\eqe
Note for future reference that $F_L$ is the same as $U(1)$ charge $e$.
The
canonical bosonization formula is\cite{stanley}
\eq
{\cal O}_{F_R,F_L}(0) = \exp \Biggl\{i \frac{F_R + F_L}{2} X^3(0) +2 \pi i
(F_R - F_L) \int_0^\infty d\sigma \pi_3(\sigma) \Biggr\}.
\eqe
The canonical momentum is
\eqa
\pi_3 &=& \frac{Q_-}{4\pi Q_+} \Bigl( \del_\zb X^3 -
\del_z X^3 + 4Q_+ A^M_\mu
\del_z X^\mu \Bigr)  \nonumber\\[2pt]
&=& \frac{Q_-}{4\pi } \del_\sigma (\xi_L - \xi_R),
\eqae
where $\xi_{R,L}$ are as defined at the end of the previous section.
The bosonization formula is then
\eq
{\cal O}_{F_R,F_L} = \exp \Bigl\{ i (F_L Q_L + F_R Q_R) \xi_L +
i (F_L Q_R + F_R Q_L) \xi_R \Bigr\}.
\eqe
In particular,
\eqa
\lambda_R &=& \exp\Bigl\{ i (Q_R \xi_L + Q_L \xi_R) \Bigr\}
\nonumber\\[2pt]
\lambda_L &=& \exp\Bigl\{ i (Q_L \xi_L + Q_R \xi_R) \Bigr\}.
\eqae
This is the same form that holds without the gauge interaction of the
fermions (but with the Thirring coupling~\ee{gthir}), though in the present
case $\xi_L$ is not a free field.

We now wish to relate the vertex operators of the monopole theory into
those of the exact CFT.  A vertex operator is a polynomial in fields and
derivatives, times a wavefunction of the embedding coordinates $X^\mu$.
There is an important subtlety: the charged fields are not globally defined.
For example, from
eq.~\ee{lampatch} we see that at the south pole $\lambda^{(N)}_{R,L}$ is
$e^{2i Q_{R,L} \phi}$ times a single-valued function.  In order to make a
conformally invariant vertex operator, we need a compensating singularity
in the wavefunction.

The necessary wavefunctions are the {\it monopole harmonics},
which we briefly
review\cite{WuYa}.
We start by recalling the relation between the ordinary spherical
harmonics and the representation matrices of $SU(2)$.  Consider first a
particle moving on the $SU(2)$ group manifold.
The representation matrices
\eq
D^j_{m_L, m_R}(g) \e{dlmm}
\eqe
form a complete set of wavefunctions on $SU(2)$.
These transform as spin-$j$ under both left- and right-multiplication,
\eq
D^j_{m_L, m_R}(g_L^{-1} g g_R) = D^j_{m_L, m'_L}(g^{-1}_L) D^j_{m'_L,
m'_R}(g) D^j_{m'_R, m_R}(g_R). \e{dtrans}
\eqe
Now let us consider the two-sphere $S_2$ regarded as a coset
$SU(2)/U(1)_R$. We obtain a complete set of wavefunctions by restricting
the set~\ee{dlmm} to those which are invariant under the
identification---namely, those with $m_R = 0$.  Left $SU(2)$ takes this
set into itself, and is the rotational symmetry of the two-sphere.  The
transformation law~\ee{dtrans} thus identifies $D^j_{m,0}(g)$ as the
spherical harmonic $Y^j_m$, up to  normalization.

Now we consider $m_R = q/2 \neq 0$.  The representation matrix~\ee{dlmm}
is no longer well-defined on the coset $S_2$, so let us make a
convenient choice of
of map $S_2 \to SU(2)$,
\eq
g_{\theta,\phi} = e^{i\phi \sigma_3/2} e^{i \theta \sigma_2/2} e^{-i \phi
\sigma_3/2}.
\eqe
With this choice, the elements
\eq
D^j_{m,q/2} (g_{\theta,\phi})
\eqe
are well-behaved at the north pole but not at the south pole.  Note that
the $\phi$-dependence of
$D^j_{m,q/2} (g_{\theta,\phi})$ is $e^{i (m - q/2) \phi}$.  At $\theta=0$,
the only nonvanishing matrix element is $m = q/2$, which is single-valued.
At $\theta = \pi$, however, the only nonvanishing element is $m = -q/2$,
which thus is multi-valued as $e^{- i q \phi}$.  Thus, $D^j_{m,q/2}
(g_{\theta,\phi})$ is the wavefunction of a particle of unit charge
in a monopole field of strength $-q/2$.

Consider also the rotational properties of these functions.  Expanding
in Euler angles, we can write
\eq
g_L^{-1} g_{\theta,\phi} = g_{\theta',\phi'} e^{i \xi' \sigma_3/2}
\eqe
for some functions $\theta'(\theta,\phi,g_L)$, $\phi'(\theta,\phi,g_L)$,
$\xi'(\theta,\phi,g_L)$.  This defines the action of $SU(2)_L$ on
the coset $S_2$,
$(\theta,\phi) \to (\theta',\phi')$.  Then
the transformation~\ee{dtrans} becomes
\eq
D^j_{m,q/2}(g_{\theta',\phi'}) =
D^j_{m,q/2}(g_L^{-1} g_{\theta,\phi} e^{-i \xi' \sigma_3/2}) = D^j_{m,
m'}(g^{-1}_L) D^j_{m',q/2}(g_{\theta,\phi}) e^{-i \xi' q/2}. \e{montrans}
\eqe
This is the spin-$j$ transformation up to a phase.  The phase, which is
independent of the state~$j,m$, is just the gauge transformation needed
to bring the Dirac string back to the south pole after a rotation.  Notice,
by the way, that there is a
lower bound on the angular momentum
\eq
j \geq \frac{q}{2}.
\eqe
Let us also note the property
\eq
-\nabla^2_{-q} D^j_{m, q/2}(g_{\theta,\phi}) = \Bigl\{ j(j+1) - q^2/4
\Bigr\} D^j_{m, q/2}(g_{\theta,\phi}),  \e{laplace}
\eqe
where $\nabla^2_{-q}$ is the covariant Laplacian for a particle of
unit charge in a monopole field $-q/2$.

We now make vertex operators by combining ${\cal O}_{F_R,F_L}$ with an
appropriate wavefunction.  Under the gauge symmetry~\ee{eplus},
\eq
\delta {\cal O}_{F_R,F_L} = i (F_L \epsilon^L + F_R \epsilon^R)
{\cal O}_{F_R,F_L}.
\eqe
The total monopole
field felt by ${\cal O}_{F_R,F_L}$ is then
\eq
q = 2 F_L Q_L + 2 F_R Q_R.
\eqe
The proper vertex operators are therefore
\eqa
D^j_{m,q/2} (g_{\theta,\phi}) {\cal O}_{F_R,F_L}
&=& D^j_{m,q/2} (g_{\theta,\phi}) \exp \Bigl\{ i q \xi/2 +
i (F_R - F_L) Q_- \xi_R \Bigr\}  \nonumber\\[2pt]
&=& D^j_{m,q/2} (g) \exp \Bigl\{ i (F_R - F_L) Q_- \xi_R \Bigr\}.
\e{boson}
\eqae
The translation between the wavefunctions~\ee{dlmm} on $SU(2)$ and
the current algebra primary fields is known\cite{gepwit},
\eq
D^j_{m_L,m_R} (g(z,\zb)) = \tilde\phi^j_{m_L}(\zb) \phi^j_{m_R}(z),
\eqe
so the final result is
\eq
D^j_{m,q/2} (g_{\theta,\phi}) {\cal O}_{F_R,F_L}(z,\zb)
= \tilde\phi^j_{m}(\zb) \phi^j_{q/2}(z) \exp
\Bigl\{ i (F_R - F_L) Q_- \xi_R(z) \Bigr\}.
\eqe
This is a satisfying result: consistent with the identification~\ee{coset}
of the theory as a coset, the vertex operators are current algebra fields
with a twist of $U(1)_R$.  Also, separating the analytic primary into
a parafermionic primary and a free boson part,
\eq
\phi^j_{m}(z) = \psi^j_m(z) e^{i m \xi_R(z)},
\eqe
we have
\eq
D^j_{m,q/2} (g_{\theta,\phi}) {\cal O}_{F_R,F_L}(z,\zb)
= \tilde\phi^j_{m}(\zb) \psi^j_{q/2}(z) \exp \Bigl\{ i (F_L Q_R + F_R Q_L)
\xi_R(z) \Bigr\}.  \e{final}
\eqe

The energy momentum tensor is
\eq
T_{zz} = \frac{1}{2(k+2)} j_z^a j_z^a, \qquad T_{\zb\zb} =
\frac{1}{2(k+2)} j_\zb^a j_\zb^a.
\eqe
We can also decompose into a parafermion part and a free boson part,
\eq
T_{zz}(z) = T^{SU(2)/U(1)}_{zz}(z) - \frac{k}{4} \del_z \xi \del_z \xi.
\eqe
The states~\ee{final} thus have weights
\eqa
\tilde L_0 &=& \frac{j(j+1)}{k+2} \nonumber\\[8pt]
L_0 &=& \frac{j(j+1)}{k+2} - \frac{q^2}{4k} + \frac{(F_RQ_L+F_LQ_R)^2}{k}
\nonumber\\
&=& \tilde L_0 + \frac{F_R^2-F_L^2}{2}.
\eqae

Let us check the correspondence with the semiclassical expansion
for some low-lying states.  The tachyon has $F_R = F_L = 0$, so the vertex
operators are just the spherical harmonics
$D^j_{m,0} (g)$.  Using the one loop result $k = 2 r_2^2$, the dimension
is
\eq
L_0 + \tilde L_0 = \frac{j(j+1)}{r_2^2 + 1},
\eqe
while the spin $L_0 - \tilde L_0 = 0$.
This matches the semiclassical value for $r_2 >> 1$, the `$+1$' being
a correction to the semiclassical result.
The vertex operator with one $\lambda_R$ oscillator excited is
\eq
D^j_{m,Q_R} (g_{\theta,\phi}) \lambda_R
= \tilde\phi^j_{m}(\zb) \phi^j_{Q_R}(z) \exp \Bigl\{ i Q_- \xi_R(z)
\Bigr\}.
\eqe
The spin is $L_0 - \tilde L_0 = \frac{1}{2}$, while the dimension is
\eqa
L_0 + \tilde L_0 &=& \frac{j(j+1)}{r_2^2 + 1} + \frac{1}{2}
\nonumber\\[2pt]
&=& \frac{j(j+1) - Q_R^2}{r_2^2 + 1}  + \frac{1}{2} + \frac{h^2}{1 + 2h}
\frac{r_2^2}{r_2^2 + 1}.
\eqae
In the last line, the first term is from the monopole
Laplacian~\ee{laplace}, while the last is the shift in the dimension of
$\lambda$ from the Thirring interaction (we have used eq.~\ee{gthir}
for the Thirring coupling).  Again the ``$+1$" represents corrections to the
semiclassical result.  Similarly, the vertex operator with one $\lambda_L$
is
\eq
D^j_{m,Q_L} (g_{\theta,\phi}) \lambda_L
= \tilde\phi^j_{m}(\zb) \phi^j_{Q_L}(z) \exp \Bigl\{- i Q_- \xi_R(z)
\Bigr\}.
\eqe
The spin is $L_0 - \tilde L_0 = -\frac{1}{2}$, and the dimension is
\eqa
L_0 + \tilde L_0 &=& \frac{j(j+1)}{r_2^2 + 1} - \frac{1}{2}
\nonumber\\[2pt]
&=& \frac{j(j+1) - Q_L^2}{r_2^2 + 1}  +
\frac{1}{2}\frac{r_2^2 - 1}{r_2^2 + 1} + \frac{h^2}{1 + 2h}
\frac{r_2^2}{r_2^2 + 1},
\eqae
with the semiclassical limit again evident.

\section*{5.\ \ Level Matching}
\refstepcounter{chapter}

Now let us make a new beginning, from a different point of view.
We start with a level $k$ $SU(2)$ WZW theory, and without reference to
any spacetime interpretation we try to make a consistent string theory by
twisting on $Z(N)_R$ for some integer $N$:
\eq
\frac{SU(2)_L \times SU(2)_R}{Z(N)_R}.
\eqe
A general twisted vertex operator is of the form
\eq
(\tilde\jmath_{{\tilde K}} \cdot \tilde \phi^j_{m_L})( j_{K} \cdot
\phi^j_{m_R}) e^{i w_3 k \xi_R / N}.   \e{twspec}
\eqe
Here $w_3$, which runs from 0 to $N-1$, is the winding number
associated to the right moving $U(1)$ boson $\xi_R$.  That is, $\xi_R$
shifts by $4\pi w_3 / N$ as $z$ encircles 0 :
\eq
\xi_R(z) e^{i w_3 k \xi_R(0) / N} \sim -\frac{2 i w_3 \ln z}{N}
e^{i w_3 k \xi_R(0) / N}.
\eqe
The notations
$j_{K}$ and $\tilde\jmath_{\tilde K}$ stand for generic products of
raising operators $j^a_{-n}$ and $\tilde\jmath_{-n}^a$
respectively.\footnote {In eq.~\ee{twspec} we have acted first with
the raising operators and then twisted.  Alternately, we can first
twist the primary, and then act with the twisted algebra of $j^3_{-n}$,
$j^\pm_{-n \mp 2w/N}$, and $\tilde\jmath_{-n}^a$.  These descriptions are
equivalent.}
The untwisted part of~\ee{twspec} depends on $\xi_R$ as $e^{i J_3\xi_R}$
and the full operator as $e^{i J'_3\xi_R}$,
where
\eq
J_3 = m_R + N_+ - N_-,\qquad
J_3' = J_3 + \frac{w_3 k}{N},
\eqe
$N_\pm$ being the number of $j^\pm$ raising operators.
The weights are therefore
\eqa
L_0 &=& \frac{j(j+1)}{k+2} + |K| + \frac{J_3^{'2} - J_3^2}{k}
\nonumber\\[2pt]
\tilde L_0 &=& \frac{j(j+1)}{k+2} + |{\tilde K}|,  \e{weights}
\eqae
where $|{\tilde K}|$, $|K|$ are the total levels of the raising operators.

A consistent string theory requires level matching.  The level mismatch
here is
\eqa
L_0 - \tilde L_0 &=& \frac{J_3^{'2}}{k} - \frac{J_3^2}{k} \quad {\rm mod}
\ {\bf Z}
\nonumber\\[2pt]
&=& \frac{w_3}{N} \Biggl( 2J_3 + \frac{w_3k}{N} \Biggr)  \quad {\rm mod}
\ {\bf Z}.
\eqae
Let us for the present section think of this CFT as a background in bosonic
string theory.  The CFT is then modular invariant by itself, so in
particular the spectrum must be restricted to states for which $L_0 -
\tilde L_0$ is an integer.  Let us focus on $w_3 = 1$; by assumption there
are states with $w_3 = 1$, or we would redefine $N \to N/(w_3)_{\rm min}$.
Then
\eq
2 J_3 + \frac{k}{N} \in N {\bf Z}.
\eqe
Since $2 J_3$ is an integer, we learn that $k/N$ must be an integer,
$N'$.  This is the quantization of the
torsion on the coset manifold.  The level matching condition is then
\eq
2 J_3 + N' \in N {\bf Z}, \qquad (w_3=1),
\eqe
which has solutions.  Since $J_3$ and $w_3$ are both conserved, the
$w_3=1$ result plus closure of the OPE implies that at general $w_3$
\eq
2 J_3 + w_3 N' \in N {\bf Z},   \e{lev1}
\eqe
or equivalently
\eq
J_3 + J'_3 \in N {\bf Z},\qquad J_3 - J'_3 \in N' {\bf Z} \e{level}
\eqe
These states satisfy level matching, and a check of the characters shows
that the partition function is in fact modular-invariant.

Now let us relate these theories to the monopole background.
Comparing the twisting and the level gives
\eq
N = 2 Q_+, \qquad N' = Q_-.  \e{nnpr}
\eqe
Also, comparing the $\xi_R$ dependences of the vertex
operators~\ee{final} and~\ee{twspec}, we have
\eq
F_R + F_L = 2\frac{J_3 + J_3'}{N}, \qquad F_R - F_L =  \frac{J'_3 -
J_3}{N'}. \e{flfr}
\eqe
The condition~\ee{level} is thus that the total right plus left fermion
number be even, which is the diagonal GSO projection.  Note
that the modular-invariant spectrum necessarily includes the Ramond-Ramond
sectors with half-integer $F_{R,L}$.

Notice that eq.~\ee{nnpr} implies that $Q_-$ is an integer, as
we found in the previous section, but allows $Q_+$ to be half-integer.  In
this case
\eq
Q_L = \frac{N + 2N'}{4}, \qquad Q_R = \frac{N - 2N'}{4}
\eqe
are not half-integer, and the
fields $\lambda_{L,R}$ have gauge-invariant Dirac string singularities.
There is a simple reason why these theories are consistent: the actual
states appearing in the GSO-projected spectrum all have properly
quantized charge.
Notice also that when $N$ is even the interchange
\eq
N \leftrightarrow 2N'
\eqe
just flips the sign of $Q_R$.  This is equivalent to the original
theory, with redefinition $\lambda_R \to \bar\lambda_R$.  In the
bosonized form this is a duality transformation on $\xi$.

This completes our construction of the monopole CFT for bosonic string
theory.  We conclude with one general remark.  Twisting by $Z(N)_R$,
one might naively restrict the spectrum to $2J'_3 \in N {\bf Z}$.
This would differ from the actual spectrum~\ee{level} by a term
proportional to winding number.  It is well known that in twisting
(orbifolding), it is not in general possible to determine the
projection naively; rather, one must explicitly solve the
level-matching condition.  We would like to point out that this is
a special case of a well-known ambiguity in field theory: one always
is free to add to a Noether current a term which is trivially
conserved, $j^a \to j^a + \del_b K^{ab}$ for any antisymmetric
$K^{ab}$, as we have done in going from the current~\ee{curr1} to
the current~\ee{curr2}.  In a topologically nontrivial sector, this
changes the conserved charge.  In the present case, there is a
simple interpretation for the actual projection.  The charge $J'_3$
corresponds (up to normalization) to the chiral current~\ee{curr2},
which one can think of as generating
\eq
\delta \xi_R = \epsilon, \qquad \delta \xi_L = 0,
\eqe
while $\frac{1}{2} (J_3 + J'_3)$ corresponds to the naive Noether
current~\ee{curr1} and generates
\eq
\delta \xi_R = \frac{\epsilon}{2}, \qquad \delta \xi_L
= \frac{\epsilon}{2}.  \e{vector}
\eqe
In a non-winding sector, only the combination $\xi_R + \xi_L$
appears, and these symmetries are equivalent, but in winding sectors
they differ.  Level matching picks out states invariant under the
vector-like symmetry~\ee{vector}.

\section*{6.\ \ Heterotic String Background}
\refstepcounter{chapter}

As discussed in section~2, the heterotic string in a monopole background
of charge $Q$ is a special case of the action~\ee{fermact}, with
\eq
Q_L = Q,\qquad Q_R = 1,  \e{charges}
\eqe
and a particular value of the Thirring interaction.
Since we found that in the bosonic case the Thirring
coupling was fixed by conformal invariance, it must be that for
$Q_R = 1$ the theory is actually superconformally invariant.

Let us show this, following the general approach of the previous
section.  We will not initially assume $Q_R = 1$.  It is well-known
that a twist of the $SU(2)$ current algebra produces an $N=2$
supersymmetry algebra:
\eqa
T_F^{\pm} &=& j^{\pm} e^{\pm i (\sqrt{(k+2)/2} - 1) \xi_R}
\nonumber\\[2pt]
&=& \psi_1^{\pm} e^{\pm i (\sqrt{(k+2)/2}) \xi_R}, \e{scurr}
\eqae
with $\psi_1^\pm$ being the parafermionic currents.  The $N=2$
algebra appears because the non-anomalous combination of
$\lambda_R$ and $\lambda_L$ chiral rotations acts on the supercharge.

We must see whether these currents survive the projection.
Level matching
is now more intricate: there are four sectors, $NS_\pm$ and $R_\pm$,
on both the left- and right-moving sides, defined by the fermion numbers
\eqa
NS_{R+}:&& F_R \in 2{\bf Z} \nonumber\\[2pt]
NS_{R-}:&& F_R \in 2{\bf Z} + 1 \nonumber\\[2pt]
R_{R\pm}:&& F_R \in 2{\bf Z} \pm \frac{1}{2}  \e{sectors}
\eqae
and correspondingly for $F_L$ on the left.  Conservation
of $F_R$ and $F_L$ implies the correct fusion $NS \cdot NS = NS$,
$R \cdot R = NS$, $R \cdot NS = R$.
In order that the full world-sheet theory satisfy level matching, a
state in a sector $(F_L,F_R)$ of the monopole CFT must have
spin
\eq
L_0 - \tilde L_0 \in \frac{F_R^2 - F_L^2}{2} + {\bf Z}.
\eqe
For
example, the sectors $NS_{L+} R_{R\pm}$ have spin $\frac{1}{8}$ and the
sector $NS_{L+} NS_{R-}$ spin $\frac{1}{2}$.

Note first that in the diagonal sectors (same on right and left),
the spin is an integer.  From the previous section, the condition
for a nontrivial spectrum is $k = NN'$.  Then if we define~$F_{L,R}$ as in
eq.~\ee{flfr} we have
\eq
L_0 - \tilde L_0 = \frac{F_R^2 - F_L^2}{2} + |K| - |{\tilde K}|
\quad {\rm mod}\ {\bf Z}  \e{sulev}
\eqe
which is precisely the correct level mismatch.

Inverting~\ee{flfr} gives
\eqa
J_3 &=& Q_R F_R + Q_L F_L \nonumber\\[2pt]
J'_3 &=& Q_L F_R + Q_R F_L.  \e{invert}
\eqae
Now, the supercurrent
has $J_3 = \pm 1$ and $J'_3 = \pm \sqrt{(NN'+2)/2}$.  Taking the
product of eqs.~\ee{flfr} implies $F_R^2 - F_L^2 = 1$.
If the supercurrent is to appear in the operator algebra~\ee{sectors} it
therefore must have $F_R = \pm 1$ and $ F_L = 0$.  Then
\ee{invert} with $J_3 = \pm 1$ implies $Q_R = \pm 1$ as the condition for
the supercurrent to appear in the algebra, as expected; we are free to take
$Q_R = 1$ by $\psi_R \leftrightarrow \bar\psi_R$.

There are several special cases.  If $Q_L = 0$ we have a neutral
remnant; however, this case requires special treatment because
$Q_L < Q_R$.  If $Q_L = 1$ (or equivalently $-1$), the level $k$
vanishes and the CFT degenerates.  We will discuss these cases
separately below, after discussing some general issues for
$Q_L \geq 2$: spacetime supersymmetry, stability, the connection
to the asymptotic spacetime, and fermion zero modes.

In order to have
spacetime supersymmetry, we need a weight $(0,\frac{1}{8})$
field in the sector $F_L = 0$, $F_R = \pm \frac{1}{2}$.  This would
combine with the $x t \times {\rm internal} \times {\rm ghost}$
spin field $S \Sigma e^{-\varphi/2}$ of weight
$(0,\frac{7}{8})$ to produce the $(0,1)$ current corresponding to
spacetime supersymmetry.
This field can only be $e^{i \xi_R \sqrt{k/8}}$,
with $J_3 = 0$ and $J_3' = \sqrt{k/8}$. Inserting into~\ee{flfr}
gives
\eq
F_R + F_L = \frac{1}{2} \sqrt{\frac{Q_-}{Q_+}},
\eqe
which is never a half-integer.
Thus these theories are not spacetime
supersymmetric, in agreement with the result
of the $\alpha'$ expansion\cite{GHS}.

Without spacetime
supersymmetry, there is the possibility of tachyons.
To get the mass-shell condition,
we need to look at the $xt$ CFT.  In the throat limit the $xt$
energy-momentum tensor is
\eq
\frac{1}{2}\Bigl( \del_z t \del_z t + \psi_t \del_z \psi_t \Bigr)  -
\frac{1}{2}\Bigl( \del_z x \del_z x + \psi_x \del_z \psi_x \Bigr) - \alpha
\del_z^2 x.
\eqe
The total four-dimensional central charge is
\eq
3 + 12\alpha^2 + \frac{3k}{k+2} = 6,
\eqe
so
\eq
\alpha = \frac{1}{\sqrt{2k + 4}} = \frac{1}{2 Q_L}.
\eqe
The primary field
\eq
e^{-\alpha x} e^{i k \cdot Y}
\eqe
with $Y^\mu = t, x$ has weight
\eq
L_0 = \frac{k_\mu k^\mu}{2} + \frac{\alpha^2}{2}.  \e{lweight}
\eqe

A scalar in the $-1$ picture has total
weights $(1, \frac{1}{2})$.
Using eqs.~\ee{weights} and~\ee{lweight}, the mass shell condition
is
\eqa
\frac{1}{2} &=& \frac{k^2}{2} + \frac{(2j + 1)^2}{8 Q_L^2}
+ \frac{F_R^2 - F_L^2}{2} + |K| + L'_0,\nonumber\\
1 &=& \frac{k^2}{2} + \frac{(2j + 1)^2}{8 Q_L^2}
+ |{\tilde K}| +\tilde L'_0,   \e{mass}
\eqae
where $L'_0$, ${\tilde L}'_0$  are the weights
from the internal CFT, plus the contribution of
any excitations from the $xt$ theory.  To have a solution with $k^2 > 0$,
it is necessary that $|K| = 0$.  The quantum number $J_3$ then comes
entirely from the primary field, so in particular
\eq
2j + 1 \geq 2|J_3| + 1 = 2|Q_R F_R + Q_L F_L| + 1.
\eqe
The only possibilities for a tachyon are then\footnote{Of course,
the could also be the unit operator from the sector $NS_+$ paired
with a state from the sector $NS_-$ of the internal theory.  Such a state,
if tachyonic, would also be tachyonic in flat space.  Since the theory is
supersymmetric in flat space, this is not possible, and we can restrict
attention to the sector $NS_-$ of the monopole theory.}
\eqa
F_R = - 1,\quad F_L = 1, && j = Q_L - 1, \nonumber\\[2pt]
F_R = - 1,\quad F_L = \frac{1}{2}, && j = \frac{1}{2}Q_L - 1.
\e{inst}
\eqae
For these,
\eq
\mu^2 = - k^2 = -\frac{|F_L| }{Q_L} + \frac{1}{4Q_L^2} + 2L'_0.
\e{masssq} \eqe
The operators are
\eq
e^{-\alpha x} e^{i k \cdot Y}
\tilde \phi^j_m \phi^j_j e^{-2i j \xi_R} {\cal V}_{\rm int}
\eqe
for the appropriate $j$ value~\ee{inst}.
We can use the bosonization formulae from section~4 to to express
the result in terms of the original Fermi fields, giving respectively
\eqa
e^{-\alpha x} e^{i k \cdot Y} D^j_{m,j} (g_{\theta,\phi}) \lambda_L \bar
\lambda_R {\cal V}_{\rm int} \nonumber\\
e^{-\alpha x} e^{i k \cdot Y} D^j_{m,j} (g_{\theta,\phi}) R_{L+} \bar
\lambda_R {\cal V}_{\rm int}.
\eqae
The internal part of the vertex operator must have spin $-\frac{1}{2}$
or $-\frac{7}{8}$ respectively
in order that the whole operator be the correct $(1,\frac{1}{2})$.
The simplest case would be a $(\frac{1}{2},0)$ field $\lambda^A_L$ from the
$NS$ sector of the current algebra or a $(\frac{7}{8},0)$ field
$R^B_{L}$ from the
Ramond sector, for which the
masses-squared~\ee{masssq} will indeed be negative.  These vertex
operators correspond to gauge bosons, and the current $\lambda^A_L
\lambda_L$ and $R^B_L R_{L+}$ are charged under the $U(1)$ of the
monopole.  In other words, these states are only present if the $U(1)$ is
embedded in an unbroken non-Abelian group in the four-dimensional gauge
group.
This instability is precisely the Brandt-Neri
instability\cite{brandtneri} to emission of non-abelian radiation; in
field theory language it arises due to the negative centrifugal potential
at $j=Q_L F_L -1$.  Even in a theory in which the $U(1)$ is embedded in a
broken non-Abelian group the black hole is destabilized if the radius
$r_2$ of the black-hole throat is small as compared to
the breaking scale.  The unstable mode produces a non-Abelian
monopole core outside the black hole horizon.  This is similar to the
instability previously studied by Lee, Nair, and Weinberg\cite{LNW}.

Aside from the non-abelian instabilities we have therefore demonstrated the
stability of the horizon and throat regions of the magnetic black holes
under linearized perturbations.

The vertex operators corresponding to changes in the radii $r_1$ and $r_2$
are both of the form
\eq
(\tilde \jmath^a_{-1} \cdot \tilde\phi^1_a)(j^b_{-1} \cdot \phi^1_c).
\e{widen}
\eqe
To see this, translate into the WZW fields\cite{WZW,gepwit}
\eq
{\rm tr}(g^{-1} \sigma^a \del_\zb g) {\rm tr}( \sigma^b g^{-1} \del_z g)
{\rm tr}(g^{-1} \sigma^a g \sigma^c ) =
2{\rm tr}(\sigma^b g^{-1} \del_z g)
{\rm tr}(g^{-1} \del_\zb g \sigma^c).
\eqe
Note in particular that this preserves rotational invariance, $SU(2)_L$.
Now, setting $b = c$ and summing yields ${\rm tr}(g^{-1} \del_z g g^{-1}
\del_\zb g)$ giving an overall change in the radius of the WZW
model, $\delta r_1/r_1 = \delta r_2 /r_2$.  Setting $b = c = 3$ changes
only the radius $r_1$.  In a bosonic theory, these are both allowed
vertex operators, but in the supersymmetric case we know that
supersymmetry relates $r_1$ and $r_2$, so only one linear combination
can be the highest operator in a superfield.  The relevant superconformal
primary is
\eq
(\tilde \jmath^a_{-1} \cdot \tilde\phi^1_a) (\phi^1_+ e^{i (Q_L - 1) \xi_R
} + \phi_- e^{-i (Q_L - 1) \xi_R }).
\eqe
Taking the operator product with the supercurrent~\ee{scurr} yields
the 0-picture vertex operator
\eq
(\tilde \jmath^a_{-1} \cdot \tilde\phi^1_a)(j^+_{-1} \cdot \phi^1_-
+ j^-_{-1} \cdot \phi^1_+ ).
\eqe
The primary is irrelevant, has weight
\eq
\Biggl(  1 + \frac{1}{Q_L^{2}}, \frac{1}{2} + \frac{1}{Q_L^{2}}
\Biggr),
\eqe
so we can make an on-shell primary by multiplying by $e^{x/Q_L}$.
This perturbation represents the widening of the throat as $x \to \infty$,
but we do not know of an exact CFT which interpolates between the throat
and asymptotic region.

Finally, we look for fermion zero modes on the throat.\footnote
{We thank Tom Banks for suggesting this.}
This gives a useful check on the construction,
and will aid us in
understanding the physics of the $|Q_L| = 1$ solution.
A massless
fermion of charge $e$ has $2eQ$ zero modes,
by the index theorem.  The state
\eq
R_{R\pm} \lambda_L S_{\pm} {\cal U}_\pm e^{- \varphi/2},
\eqe
is a massless fermion in four dimensions, where $S_{\pm}$ is the
$(0,\frac{1}{8})$ spin field from the $xt$ theory, $\cal U_\pm$
is a $(\frac{1}{2}, \frac{3}{8})$ field from the internal theory,
$\varphi$ is the bosonized ghost, and $R_{R\pm}$ are the Ramond ground
states from the monopole CFT.
Its $U(1)$ charge is $e=1$ from
$\lambda_L$.  The appropriate
spatial wavefunction is $e^{-\alpha x} e^{i k \cdot Y}
D^j_{m,Q_L \pm 1/2} (g_{\theta,\phi})$.  The mass shell condition
$L_0 = 1$ is then
\eq
-k^2 = \frac{(2j+1)^2}{4Q_L^2} - F_L^2  \e{fermass}
\eqe
with $F_L$ = 1.
Recalling that $j \geq q/2$, the lowest state is $R_{R-} \lambda_L$
at $j = Q_L - \frac{1}{2}$.  This is indeed massless, and its
multiplicity is $2j+1 = 2Q_L$ as expected,
This can also be phrased as an
index.  The spectrum is
\eqa
R_{R+}:&&j = \qquad\quad Q_L + \frac{1}{2}, \quad Q_L + \frac{3}{2},
\quad\ldots
\nonumber\\
R_{R-}:&&j =
Q_L - \frac{1}{2}, Q_L + \frac{1}{2},
Q_L + \frac{3}{2},
\ldots.
\eqae
The Ramond generator $G_0$ takes $R_{R+} \leftrightarrow R_{R-}$
and anticommutes with $(-1)^{F_R}$.  We see that all the massive states
are appropriately paired, while the massless states are the
kernel of $G_0$.  Another example is
\eq
R_{R\pm} R_{L+} S_{\pm} {\cal U}'_\pm e^{- \varphi/2},
\eqe
which for ${\cal U}'_\pm$ of weight $(\frac{7}{8},\frac{3}{8})$ is
a massless fermion of charge $e = \frac{1}{2}$.  The mass
shell condition is again~\ee{fermass}, with $F_L = \frac{1}{2}$,
and $R_{R-} R_{L+}$ at $j = \frac{1}{2} Q_L - \frac{1}{2}$
indeed gives a massless
state of multiplicity $Q_L$.

\section*{7.\ \ The Neutral Remnant: $Q=0$}
\refstepcounter{chapter}

A vanishing monopole field corresponds to $Q_R =1$, $Q_L = 0$.
Since our construction assumes $Q_R < Q_L$, we need to take
the case $Q_L = 1$, $Q_R = 0$ and flip:
\eq
\frac{SU(2)_L \times SU(2)_R}{Z(2)_L}.
\eqe
It might seem strange that there is a neutral solution, since the
idea was to balance curvature against magnetic field.  In the
bosonic form, the curvature is balanced by torsion which arises
from the spin connection, both one-loop effects.  In the original fermionic
variables, a one-loop curvature term is balanced by a two-loop
curvature-squared term.  Obviously, we cannot be sure from the semiclassical
picture that this is possible, but the exact construction shows that it is.
The WZW model is known to be conformally invariant, and the $Z(2)$ modding
does not affect the $\beta$-functions.

The vertex operators are then the parity transform of the previous
discussion,
\eq
e^{i w_3 \zeta_L} ( \tilde\jmath_{\tilde K} \cdot \tilde\phi^j_{m_L} )
(j_{K} \cdot
\phi^j_{m_R} ) .  \e{ops}
\eqe
Here, $\zeta_L$ is the antianalytic field obtained by bosonizing
$U(1)_L$, not the same as the earlier $\xi_L$
associated with $U(1)_R$.  The fermion numbers are
\eqa
\tilde J_3 &=& F_R  \nonumber\\[2pt]
\tilde J'_3 &=& F_L,
\eqae
where $\tilde J_3$ is the left $J_3$ value of the untwisted
operator, and $\tilde J'_3 = J_3 + w_3$ is the left $J_3$ value of the
twisted operator.  The spectrum runs over all operators of the form~\ee{ops}
with half-integer $F_R$ and $F_L$.

The $k=2$ current algebra can be represented by a triplet of current
algebra fermions, $\psi^a_{R,L}$.  Note that the original $\lambda_L$ are
free, because they do not couple to the gauge field, and one finds
the simple result
\eq
\lambda_L = e^{i \zeta_L} = \psi^1_L + i\psi^2_L = \tilde\phi^1_+ .
\eqe

Let us check the supercurrent.  Taking $\tilde J_3 = \pm1$, $\tilde J_3' =
0$,
one obtains the $(0, \frac{3}{2})$ operators
\eq
e^{\mp i \zeta_L} \tilde\phi^1_{\pm} j^a_{-1} \cdot \phi^1_m.
\eqe
In fact, there is only one independent operator here.  The left-moving side
is the unit operator, as it must be since it has zero weight.  The right-hand
side is seen from the free fermi representation to be
$\delta^a_m$ times
\eq
\psi_R^1 \psi_R^2 \psi_R^3
\eqe
the well-known fermionic supercurrent.  In this case there is no
promotion to $N = 2$: because $Q_L = 0$, the non-anomalous $U(1)$
acts only on $\lambda_L$ and commutes with the supercurrent.

There are no left-moving supersymmetries.  Candidates such as
$e^{\pm i \zeta_L (\sqrt{2} - 1)} \tilde \jmath^\pm$
and
$\psi_L^1 \psi_L^2 \psi_L^3$
are not of the form~\ee{ops}.  The $SU(2)$ rotational symmetry is now
on the right-moving side.  There is no $SU(2)_L$ symmetry.  One would
have expected
this to survive a $Z(2)$ projection, since this is in the center, but
the GSO projection uses a $Z(4)$ twist and eliminates the $SU(2)_L$
generators.

Again there is no spacetime supersymmetry: $L_0 = j(j+1)/4$
cannot take the value $\frac{1}{8}$.  Nor can it take the value
$\frac{1}{16}$, as would be needed for a fermion zero mode;
this confirms the identification $Q = 0$.  The solution is stable: other than
the tachyon, the lowest weight in the $NS_R$ spectrum is $\frac{1}{2}$.

There is a technical subtlety, though it
presumably does not affect the physics in the
end.  We believe that this neutral remnant can join smoothly to an
external asymptotically flat spacetime, although we cannot construct
the exact solution.  We should expect, however, to be able to identify the
operators in the CFT corresponding to the perturbative broadening of the
throat, as in the previous section.  The vertex operator~\ee{widen} from the
previous section reduces for $k=2$ to
\eq
\delta^b_c \psi^1_L \psi^2_L \psi^3_L \psi^1_R \psi^2_R \psi^3_R.
\eqe
There is only one independent operator, whereas we have two radii.  Moreover,
this is not the operator we want.  It is not the top state in a world-sheet
superfield.
Also, it preserves the full $SU(2)_L \times SU(2)_R$ symmetry of the action,
whereas the supersymmetric deformation affects only $r_2$, since the
Thirring interaction vanishes when $Q_L = 0$.

One might be tempted to assume that there is no vertex operator for
deformation
of $r_2$ (so that this neutral throat could not join to an asymptotic
spacetime).  This is surely not the case.  Over some finite region of
$(r_1,r_2)$ parameter space, the $(Q_L,Q_R) = (0,1)$ sigma model
must flow to the WZW fixed point.
In particular, start near near the $SU(2) \times SU(2)$ line $r_1 =
r_2$; this line is attractive, and on the line we flow to
the WZW point.  So the WZW theory must have separate moduli for deformation of
$r_1$ and $r_2$.  The problem is that at small $k$, the identification
of sigma model operators in the CFT becomes more complicated.  For example,
the wavefunction $D^j_{m,m'}(g)$ corresponds to a primary only for $k \geq
2j$.  So we need to look for the necessary vertex operator.  It has
several identifying properties: a scalar under $SU(2)_R$, a tensor under
$SU(2)_L$, independent of $\psi_L^{1,2}$ since these are always free,
and invariant under world-sheet supersymmetry.  An operator with
these properties first appears at level $(2,2)$:
\eq
\psi_L^3 \del_\zb \psi^3_L \sum_{a=1}^3 \psi^a_R \del_z \psi^a_R.
\eqe
This is globally supersymmetric (its world-sheet integral is annihilated
by $G_{-1/2}$) and becomes locally supersymmetric and $(1,1)$ with
Liouville dressing.

\section*{8.\ \ The Degenerate Remnants: $Q = \pm 1$}
\refstepcounter{chapter}

For $Q_L = Q_R = 1$, the level $k = 2Q_+ Q_-$ vanishes.  In the
bosonized form, there is no torsion due to the left-right symmetry, and so no
nontrivial fixed point.  One might conclude that there are no $Q=1$
 remnants of
this form.  However, we would like to suggest a more interesting
possibility.  Namely, that there is a throat, in which the angular CFT has
collapsed to the trivial $k = c = 0$ theory, leaving only the the $xt$
theory and the internal theory.

One way to motivate this is via the renormalization group.
Roughly speaking, we would like
to think about the radial dependence as a renormalization group flow.  For
example, in the solutions found thus far, if we have an asymptotically flat
external spacetime connected to the throat, then $r_2$ flows from $\infty$
to the WZW fixed point value.  This identification of radius with world-sheet
scale is quantitatively accurate near the WZW fixed point (i.e. in the
throat region), because the radial
``dressing" needed to convert the kinetic term for the angular
modes to a $(1,1)$ operator is near to the identity.   However near
and outside of the mouth this is no longer the case. It would be
quantitatively accurate along the
entire flow if the dilaton had a large (in string
units) radial derivative.  When the dilaton has a radial gradient,
the radial coordinate has a classical scale transformation
\eq
\delta x = \delta \epsilon G^{xx} \del_x \Phi,
\eqe
so that
the radius would approximate a classical Liouville field if the
gradient were large.  In the present case, the derivative is of order one
in string units, so we must hope that the renormalization group picture is
a reasonable qualitative guide.  Similar ideas
have recently been discussed by Polyakov\cite{poly}, in a cosmological
context, and Banks\cite{Tom}, in the context of Schwarzchild black holes.

At very large radius, the complete theory can
presumably
be analyzed in a sigma-model expansion in $\frac{1}{r_2}$.
This is equivalent to perturbatively
solving the low-energy field equations. One finds that
the geometry is asymptotically flat, and that to evolve to smaller $r_2$
one must specify as initial data the mass $M$ and the charge
$Q$.\footnote{We consider here theories for which the
coefficient of $\partial_+ t \partial_-t$ is independent of $r_2$;
this excludes non-extremal black holes, and leads to the constraint
$M^2-2MD-D^2+Q^2/2=0$.  More generally the dilaton charge $D$ is
required as additional initial data. Our normalization of D is such that the
force between two black holes is proportional to
$(M_1M_2+D_1D_2-Q_1Q_2/2)$.}  Evolving inward,
one eventually reaches the region where $r_2$ is of order one and (if
$Q$ is also order one) geometric curvatures become of order one and
sigma-model perturbation theory breaks down.  For generic values of $M$ and
$Q$ one expects to encounter a singularity inside this region -- one
certainly  hopes that there are no smooth, negative-mass solutions of
string  theory!  However by tuning $M$ relative to $Q$, it may be possible to
flow into a throat region, in which the theory may be approximately
analyzed  in terms of radial-renormalization group flows of the angular
theory  on $S_2$. This is certainly the case for large $Q$, and in the
following we assume that it is also true for $Q=1$.

The $Q = 1$ monopole theory is a nonlinear sigma model
without torsion.  It is thus expected to flow to strong
coupling and develop a mass gap,
so the low energy theory is the trivial $c=0$ CFT.  This is consistent
with studies of this particular
model\cite{cp1}, which is also (in fermionic form)
the $(2,2)$ supersymmetric $O(3)$ nonlinear sigma model.
Applying this to the black hole, the sigma model would become strongly coupled
in the mouth region, and in the throat only the $c=0$ theory would survive, as
suggested above.

Let us consider the spectrum of the string theory in the throat.  The sigma
model has a $Z(2)$ chiral symmetry
\eq
\lambda_L \to - \lambda_L, \qquad
\lambda_R \to \lambda_R
\eqe
which is non-anomalous: the instanton amplitude from $S_2 \to S_2$ with
unit winding is of the form $\lambda_L^{2Q_L} \bar \lambda_R^2$.  This $Z(2)$
is broken at strong coupling\cite{cp1},
\eq
< \lambda_L \bar \lambda_R > \propto \pm 1.  \e{vacua}
\eqe
At low energy---that is, in the throat---the characteristic
length scale of the sigma model is large compared to the size of the string.
There are therefore two ground states.  To be precise,
there are two such vacua, $\ket{\pm}_{RR}$ (eigenstates of the
order parameter ~\ee{vacua} ), in the purely periodic $R_L R_R$
sector.  Both of these are supersymmetric, as we will see explicitly below
from an index theorem.  Fermion number acts on these as
\eq
(-1)^{F_R} \ket{\pm}_{RR} = i \ket{\mp}_{RR}, \qquad
(-1)^{F_L} \ket{\pm}_{RR} = -i \ket{\mp}_{RR}.
\eqe
These are determined, up to redefinition, by the requirement that
$(-1)^{F_{R,L}}$ square to $-1$ in the Ramond sector,
and by the requirement that $(-1)^{F_R + F_L}$, being part of a
continuous $U(1)$ symmetry, leave the ground state invariant.  The fermion
number eigenstates are
\eq
\ket{+}_{RR} \pm \ket{-}_{RR}, \qquad (-1)^{F_R} = - (-1)^{F_L} = \pm
i. \e{RR}
\eqe
In the $R_L NS_R$ and $NS_L R_R$ sectors the order parameter $\lambda_L
\bar\lambda_R$ is antiperiodic.  All such states must contain a kink, and
so are massive.  In the $NS_L NS_R$ sector, there are again two ground
states, $\ket{\pm}_{NS\,NS}$, with
\eq
(-1)^F \ket{\pm}_{NS\,NS} = (-1)^{F_L} \ket{\pm}_{NS\,NS} =
\ket{\mp}_{NS\,NS}.  \e{NSNS}
\eqe
The fermion number
eigenstates are
\eq
\ket{+}_{NS\,NS} \pm \ket{-}_{NS\,NS}, \qquad (-1)^F = (-1)^{F_L} = \pm
1.
\eqe
Although supersymmetry is broken in this sector, the effects are
exponentially
small (in the ratio of the string size to the length scale of the sigma
model) due to the mass gap.
In the throat region the monopole CFT thus has the four states~\ee{RR}
and~\ee{NSNS}, all of essentially zero weight.  One usually expects only
a single zero-weight state, but these states are distinguished by the massive
degrees of freedom.

{}From the index theorem for spacetime zero modes discussed in section~6, we
expect one state of weight $(0,0)$ in each of the sectors $R_{L+} R_{R-}$
and $R_{L-} R_{R+}$. These we have found, so the throat theory is
nontrivial: there are massless fermions.  The earlier discussion would
also lead us to expect two states of weight $(\frac{3}{8},0)$ in each of
the sectors $NS_{L-} R_{R\pm}$. These cannot exist if there is a mass gap:
only zero weight is allowed in the trivial CFT.  In fact, these states get
mass from instantons.  The mass in the effective $xt$ Dirac equation for
these two states is \eq
\bra{\bar \lambda_L R_{R+}} | G_0 \ket{ \lambda_L R_{R-} }.
\eqe
Precisely for $Q_L = 1$ the instanton amplitude $\lambda_L^{2Q_L} \bar
\lambda_R^2$ allows such a matrix element to be nonvanishing.

For completeness let us state the index argument in a careful way.  The index
is
\eq
I = {\rm tr} \Bigl( \delta_{L_0 - \tilde L_0, s} e^{\pi i F_L/Q_L + i \pi
F_R} \Bigr),
\eqe
the trace being taken in the supersymmetric $R_R$ sector.
This is $(-1)^{F_R}$, restricted to a sector of given spin and weighted by
the $U(1)$ charge $F_L$ modulo the anomaly.  Note that
the operator in the trace is a non-anomalous symmetry, and anticommutes
with $G_0$.  We will calculate this in two limits.  The first is $r_2$ very
large.  Here, acting on a state whose right-moving part is $R_{R\pm}$ and
whose left-moving part has fermion number $F_L$, the Ramond generator
reduces to the Dirac operator for $e = F_L$ on the
two-sphere, and so this state should contribute $\mp i 2 Q_L |F_L|
e^{\pi i F_L/Q_L}$ to the trace. Thus, we have from the discussion in
section~6,
\eqa
s = 0:&& I = 2Q_L \sin (\pi/2Q_L) \nonumber\\[2pt]
s = -\frac{3}{8}:&& I = 4Q_L \sin(\pi /Q_L).
\eqae
For $Q_L = 1$, the second index vanishes, so there is no contradiction
with the spectrum we have found.

The theory is not spacetime supersymmetric; there are no $NS_L R_R$ states
at all in the throat.  As before, it is unstable if embedded in a non-Abelian
gauge group due to low-lying $NS_{R-}$ states, but becomes stable if
the non-Abelian bosons are sufficiently heavy, eq.~\ee{masssq} giving $m^2
\geq 1/4$.

The angular dimensions have effectively shrunk to zero radius, leaving
behind some massless states but no other angular excitations at all.\footnote
{Another curiousity: regarded as a bosonic compactification, there are
effectively {\it two} throats.  Strings in different chiral
vacua~\ee{vacua} cannot interact in the throat because the cost would be
prohibitive.  There are even two non-interacting gravitational fields,
and so the picture of two throats appears to be a natural interpretation.
In the heterotic theory, the GSO projection always forces a particular
linear combination of the two vacua.}
This is a counterexample to the phenomenon of duality familiar from tori.

\section*{9.\ \ Conclusions}
\refstepcounter{chapter}

We have studied a string theory with right-moving supersymmetric fermions and
left-moving current algebra
fermions.  There are various generalizations of our construction.
For example, supersymmetric magnetic black holes
\cite{GiHu,GiMa} are solutions of string theory
when the $U(1)$ charge arises from
toroidal compactification. At least in some
cases these can be described beginning from a
supersymmetric version of \ee{co}.  Recently there
has been some discussion\cite{harv} of non-Abelian
black holes and monopoles related to
compactifications of the symmetric fivebrane\cite{wsheets}.  These may
also be constructed with
left-right symmetric generalizations of our procedure.  Finally, by an
electromagnetic duality rotation in the effective field theory one may
construct dyon solutions\cite{STW}.  These solutions also have infinite
throats like the purely magnetic black holes.  Although these non-trivially
mix the $xt$ theory with the angular theory, one might hope to describe
them by a related construction.

Exact, modular invariant $(0,2)$ CFTs have made only rare
appearances in the literature. The examples given here might
also be used for $(0,2)$ compactifications of the heterotic string.
Indeed, one of the original motivations of this work was a
(thus far unsuccessful) search for models
with small tree-level supersymmetry breaking.

Potentially the most interesting - and certainly the most tenuous -
result of this paper is the stringy resolution of timelike gravitational
singularities by an infinitely massive two-dimensional field theory.
While only one particular
case has been discussed in detail, the idea is
clearly very general. We certainly feel that it merits further investigation.
\\[20pt]
{\Large \bf Acknowledgements}

We thank Tom Banks, Mike Douglas, Jeff Harvey, and
Steve Shenker for helpful discussions.  This work was supported by
DOE grant DOE-91ER40618 and by NSF grants PHY-91-57463, PHY-90-09850,
and PHY-89-04035.

\end{document}